\documentclass[a4paper,11pt]{article}
\usepackage{jcappub} 
\usepackage{lineno}
\usepackage{hyperref}
\usepackage{mathtools}

\def\LCDM{$\Lambda$CDM}
\def\CDM{CDM}

\def\ns{$n_{\rm{s}}$}
\def\Omk{$\Omega_{\rm{K}}$}
\def\nrun{$\mathrm{d}n_{\mathrm{s}}/\mathrm{d}\ln k$}
\def\shortnrun{$n_{\rm{run}}$}
\def\wo{$w_{\rm{0}}$}
\def\wa{$w_{\rm{a}}$}
\def\neff{$N_{\rm{eff}}$}

\arxivnumber{} 
\title{Evaluating extensions to LCDM: an application of Bayesian model averaging and selection}

\author[a,b]{S. Paradiso\note{Corresponding author.}}
\author[c]{G. McGee}
\author[a,b,d]{W.J. Percival}

\affiliation[a]{Waterloo Centre for Astrophysics, University of Waterloo,\\
Waterloo, ON N2L 3G1, Canada}
\affiliation[b]{Department of Physics and Astronomy, University of Waterloo,\\
Waterloo, ON N2L 3G1, Canada}
\affiliation[c]{Department of Statistics and Actuarial Sciences, University of Waterloo,\\
Waterloo, ON N2L 3G1, Canada}
\affiliation[d]{Perimeter Institute for Theoretical Physics,\\
31 Caroline St North, Waterloo, ON N2L 2Y5, Canada}

\emailAdd{simone.paradiso@uwaterloo.ca}

\abstract{We present a powerful and innovative statistical framework to address key cosmological questions about the universe's fundamental properties, performing Bayesian model averaging (BMA) and model selection. Utilizing this framework, we systematically explore extensions beyond the standard \LCDM\ model, considering a varying curvature density parameter \Omk, a spectral index \ns$=1$ and a varying \shortnrun, a constant dark energy equation of state (EOS) \wo\CDM\ and a time-dependent one \wo\wa\CDM. We also assess cosmological data against a varying effective number of neutrino species \neff. Our analysis incorporates data from various combinations of cosmic microwave background (CMB) data from the latest Planck PR4 analysis, CMB lensing from Planck 2018, baryonic acoustic oscillations (BAO), and the Bicep-KECK 2018 results.

We reinforce the standard \LCDM\ model statistical preference when combining CMB data with CMB lensing, BAO, and Bicep-KECK 2018 data against the K-\LCDM\ model and \nrun-\LCDM\ with a probability $> 80\%$. When evaluating the dark energy EOS, we find that this dataset does not exhibit a strong preference between the standard \LCDM\ model and the constant dark energy EOS model \wo\CDM, with a model posterior probability distribution of approximately $\approx40\%:60\%$ in favour of \wo\CDM, while the time-varying dark energy EOS model only holds below $1\%$ probability. We find a similar result also when considering the \neff-\LCDM\ model, with a split probability almost 50\%-50\% from both our datasets.

Overall, our application of BMA reveals that including model uncertainty in these cases does not significantly impact the Hubble tension, showcasing BMA's robustness and utility in cosmological model evaluation.
}

\notoc

\newcommand\MLsimeq{\stackrel{\mathclap{\normalfont\mbox{\tiny ML}}}{\simeq}}

\begin{document}
\maketitle
\flushbottom

\section{Introduction}
\label{sec:intro}

The \LCDM\ cosmological model has been the prevailing paradigm for the last 25 years, successfully encapsulating the large-scale structure and evolution of the Universe in a model with a relatively small number of parameters (\cite{Bennett:2013,planck2018-VI} and refs therein). However, as our observational capabilities have advanced, a nuanced exploration of the model's foundations has become imperative. Bayesian model averaging (BMA) and selection is a robust statistical framework for combining and comparing competing models, allowing for a comprehensive assessment of various cosmological extensions \cite{Hoeting:1999}. We investigate the applications of BMA within the context of the \LCDM\ model, employing chain post-processing methodologies as presented by \cite{Paradiso:2023} to rigorously evaluate the model posterior probabilities associated with key extensions.

We focus on four possible extensions to the \LCDM\ model: (1) the introduction of a dynamical dark energy component, parameterized by the equation-of-state $w(z)$; (2) the inclusion of spatial curvature, characterized by the parameter $k$; (3) variations in the primordial power spectrum, encapsulated by the power law index $n_s(k)$; and (4) an additional neutrino species. Each of these extensions has distinct implications for the cosmic evolution and structure formation, necessitating a careful examination of their likelihood within the observational constraints. The inclusion of spatial curvature is motivated by both theoretical considerations and observational hints, prompting a reevaluation of the universe's global geometry \cite{planck2018-VI}. Dark energy, represented by its equation of state, remains a key driver of cosmic acceleration, and probing its potential evolution introduces an additional layer of complexity \cite{Frieman:2008}. The running of the spectral index introduces a departure from the scale invariance of primordial density fluctuations, while the presence of tensor modes directly relates to the elusive gravitational wave background from cosmic inflation.

To tackle this multifaceted exploration, we adopt the chain post-processing techniques outlined in \cite{Paradiso:2023}. This provides a systematic and computationally efficient means of extracting precise model posterior probabilities from cosmological parameter chains generated through Markov Chain Monte Carlo (MCMC) simulations for individual model choices. By applying these techniques to extended \LCDM\ scenarios, we aim to obtain a nuanced understanding of the statistical evidence and viability of each model within the Bayesian framework. The four  possible extensions discussed above are tested using the following model comparisons: (1) \LCDM, and \wo\CDM\ which considers a varying equation-of-state (parameterized by \wo), and \wo\wa\CDM\ with both \wo\ and a time-varying parameter \wa; (2) \LCDM\, and $k$\LCDM\ with spatial curvature $k$; (3) \LCDM\, and \LCDM\ with a fixed spectral index---corresponding to a Harrison-Zeldovich power spectrum (hereon HZ) with \ns$=1$---and \LCDM\ with a running spectral index, commonly denoted as \shortnrun$\,=\,$\nrun; and (4), we test the \LCDM\ model against a varying number of neutrino species $N_{\rm eff}$.


In undertaking this research, we align with the growing body of literature that recognizes the importance of sophisticated statistical methods when confronting the complexities of modern cosmological data \cite{Liddle:2004,Trotta:2008,Paradiso:2023}. This work proposes a step towards refining our cosmological framework, ensuring that our understanding of the universe remains aligned with the latest observational evidence. 

As well as allowing us to assign probabilities to models, BMA also provides a mechanism for deriving credible intervals for cosmological parameters common to multiple models, when marginalising over the model choice. Thus, for example, we can show how the estimate of the Hubble parameter today, $\rm H_0$, changes according to the additional model uncertainty. For the Hubble parameter, this is particularly interesting given the "Hubble tension", where incompatible measurements have been made using different techniques. The CMB data reported a value of $H_0$ close to $68$kms$^{-1}$Mpc$^{-1}$ for \LCDM\ cosmologies \citep{planck2018-VI}. This is supported by the combination of observations of Baryon Acoustic Oscillation (BAO) from the Sloan Digital Sky Survey (SDSS) and Big Bang Nucleosynthesis (BBN) constraints, which favour similar values for the same models \citep{eboss-cosmology}. 
In contrast, observations from local distance ladder studies, such as the Hubble Space Telescope (HST) observations of Cepheid variable stars, favour a higher value of $H_0\sim74$kms$^{-1}$Mpc$^{-1}$  \citep{Riess:2019cxk}. We consider how the error on $H_0$ varies when marginalised over the extensions to the \LCDM\ model listed above. We find increased credible intervals, in line with other analyses \citep{DIVALENTINO:2016,Hill:2020,Archidiacono:2020,Dainotti:2021}, but that such extensions do not solve the problem \citep{DiValentino:2021}. The extension of BMA to Early Dark Energy models, designed to solve this tension was considered in \cite{Paradiso:2023}.


The structure of this paper is as follows: In Sect.~\ref{sec:method} we summarize the BMA methodology and implementation. We then show our results by considering different combinations of cosmological models in Sect.~\ref{sec:results}. In Sect.~\ref{sec:conclusions} we conclude with a comment on our findings and future perspectives about the application of BMA to model selection in Cosmology.

\section{Bayesian Model Averaging and Selection}
\label{sec:method}

Bayesian model averaging (BMA) and selection offers a principled approach to deal with the choice of various cosmological models. At the heart of BMA lies a departure from the conventional model selection paradigm where, rather than choosing a single \textit{best} model and then considering parameters within that model, one can instead jointly consider a set of models and obtain combined estimates for common parameters. In essence, model choice becomes a discrete parameter to be constrained or marginalised over like any other. By assigning posterior probabilities to a set of candidate models based on the observed data, BMA combines estimates from each model, weighting them according to their posterior model probabilities. This results in a comprehensive and nuanced representation of the uncertainty associated with model choice, fostering a robust and accurate inference process.

BMA differs from conventional 
manual or automatic post-hoc selection strategies based on goodness of fit measures likeAkaike Information Criterion (AIC)\citep{Akaike:1974}. 
Such methods focus on identifying a single optimal model, neglecting the fact that there is uncertainty in this choice much like uncertainty in parameter estimates (see for instance \citep{Tan:2012}). BMA, with its probabilistic framework, provides a more inclusive approach by considering a range of models, thus offering a more realistic reflection of the underlying uncertainty. Other approaches include cross-validation methods \citep{Stone:1974}; while valuable for assessing model performance, these may not explicitly address model uncertainty. BMA, on the other hand, explicitly incorporates uncertainty through probabilistic model averaging, allowing for a more nuanced understanding of the uncertainty associated with parameter estimates and predictions.


To see how BMA and the approach of \cite{Paradiso:2023} work, let $M=1,2,...,K$ be a discrete random variable indicating a choice among $K$ different models. For convenience, we will refer to $M=j$ as $M_j$. The posterior model posterior probability for model $M_{\rm{i}}$, denoted as $P(M_{\rm{i}}\vert\Vec{d})$, is computed as
\begin{equation}
    P(M_{\rm{i}}\vert\Vec{d}) = \frac{ P(M_{\rm{i}}) P(\Vec{d}\vert M_{\rm{i}} )}{ \sum_{\rm{j}=1}^{\rm{K}} P(M_{\rm{j}}) P(\Vec{d}\vert M_{\rm{j}}) }\,,
    \label{eq:BMA}
\end{equation}
where $P(M_{\rm{i}})$ is the prior probability of model $M_{\rm{i}}$, $P(\Vec{d}\vert M_{\rm{i}})$ is the marginal likelihood of the data $\Vec{d}$ given model $M_{\rm{i}}$, and the denominator serves as a normalization factor to ensure that the probabilities sum to one. We assume that the prior probabilities are separable between model selection and model parameters.
Given the model posterior probability $P(M_{\rm{i}}\vert\Vec{d})$, we can marginalize over the model uncertainty and compute the model parameters' posterior distribution as:
\begin{equation}
    P(\Vec{\theta}\vert \Vec{d}) = \sum_{i=1}^{K}P(\Vec{\theta}, M_i\vert \Vec{d}) = \sum_{i=1}^{K} P(\Vec{\theta}\vert \Vec{d},M_i)P(M_i\vert\Vec{d})\,.
    \label{eq:marginal_posterior_BMA}
\end{equation} 
 Since samples of $P(\Vec{\theta}\vert \Vec{d},M_1)$ and $P(\Vec{\theta}\vert \Vec{d},M_2)$ are readily available from standard analyses conditional on a single model, it then only remains to obtain the two model posteriors $P(M_{\rm{i}}\vert\Vec{d})$ defined in Eq.~\ref{eq:BMA}. This decomposition frames the model-averaged posterior $P(\Vec{\theta}|d)$ as the weighted average of the model-conditional posteriors $P(\Vec{\theta}|\Vec{d},M_i)$ as in Eq. \ref{eq:marginal_posterior_BMA}. Note that it is not sufficient to weight by the model priors $P(M_i)$; rather the weights correspond to the model posteriors $P(M_i|\Vec{d}),$ which can be estimated via a suitable estimator, such as the harmonic mean (see, \cite{Newton:1994,Paradiso:2023}):
 \begin{equation} 
    \hat{P}(\Vec{d}\vert M_{\rm{i}}) =\frac{\rm{N}_{\rm{i}}}{\sum_{\rm{t}=1}^{\rm{N}_{\rm{i}}}\left[ 1/P(\Vec{d}\vert M_{\rm{i}}, \Vec{\theta}_{\rm{i}}^{(\rm{t})})\right]}\,.
    \label{eq:BMA_like_estimator}
\end{equation}
As discussed in the literature, the harmonic mean may suffer from a large variance, and alternative estimators have been proposed \cite{Newton:1994}. Recently, an evolution of the harmonic mean estimator has been presented in \cite{mcewen:2022}, the \textit{learnt harmonic mean estimator}, by interpreting the harmonic mean estimator in terms of importance sampling and adding a learned target distribution $\varphi(\Vec{\theta}) \MLsimeq \varphi^{\mbox{\tiny optimal}}(\Vec{\theta})$ with the goal of minimizing the estimator variance:
\begin{equation}
    \hat{P}(\Vec{d}\vert M_{\rm{i}}) =\frac{\rm{N}_{\rm{i}}}{\sum_{\rm{t}=1}^{\rm{N}_{\rm{i}}}\left[ \varphi(\Vec{\theta})/P(\Vec{d}\vert M_{\rm{i}}, \Vec{\theta}_{\rm{i}}^{(\rm{t})})\right]} .
    \label{eq:BMA_lhm_estimator}
\end{equation}
Here  $\varphi^{\mbox{\tiny optimal}}(\Vec{\theta})$ is learned  by training a machine learning model on a fraction of the chains --typically the first half-- and the remaining is used to compute the estimator in \ref{eq:BMA_lhm_estimator} using the publicly available  \textsc{Harmonic}\footnote{\hyperlink{https://astro-informatics.github.io/harmonic/index.html}{https://astro-informatics.github.io/harmonic/index.html}}\cite{harmonic} package.

The data themselves inform how much weight is assigned to either model. We therefore applied BMA by importance sampling existing model conditional chains \citep{Hastie:2012}. This methodology has been implemented for cosmological analyses in \cite{Paradiso:2023}, and referred as \texttt{Fast-MPC}. Details of the  implementation can be found in \cite{Paradiso:2023}. Briefly, one can apply  BMA to  existing chains as follows:
\begin{enumerate}
    \item Fit candidate models: run two or more MCMC chains assuming model $M_i$, yielding $\{\Vec{\theta}_i\}_{\rm{N}_i}$, $\{P(\Vec{d}\vert M_{\rm{i}}, \Vec{\theta}_{\rm{i}})\}_{\rm{N}_i}$.
    \item Estimate ${P}(\Vec{d}\vert M_{i})$: compute estimates $\hat{P}(\Vec{d}\vert M_{i})$ via Eq.~\ref{eq:BMA_like_estimator} or Eq.~\ref{eq:BMA_lhm_estimator}.
    \item Estimate model posteriors: substitute estimates $\hat{P}(\Vec{d}\vert M_{i})$ into Eq.~\ref{eq:BMA} to estimate $P(M_i|d)$.
    \item Estimate marginal (model-averaged) posteriors of $\Vec{\theta}$: substitute estimates of $P(M_i|d)$ into Eq.~\ref{eq:marginal_posterior_BMA}.
\end{enumerate}

\section{Model choices in cosmology}
\label{sec:results}

For our analysis, we considered the following combination of \LCDM\ base parameters with flat prior ranges, unless otherwise specified: the present day value of the Hubble parameter $\rm H_0\in[20,100]$, the baryon density parameter $\Omega_{\rm{b}}\rm{h}^2\in[0.005,0.1]$, the cold dark matter density parameter $\Omega_{\rm{c}}\rm{h}^2\in[0.001,0.99]$, the amplitude of scalar perturbations $\log(10^{10} A_\mathrm{s})\in[1.61,3.91]$, the index of the scalar perturbations $n_{\rm{s}}\in[0.8,1.2]$ and the optical depth of reionization $\tau\in[0.01,0.8]$. In addition, we include the following \LCDM\ extensions when considering the respective extended \LCDM\ model: the curvature density parameter $\Omega_{\rm k}\in[-0.3,0.3]$, the DE EOS parameters \wo$\,\in[-3,0.3]$ and \wa$\,\in[-3,2]$, the running of the primordial fluctuations spectral index \shortnrun$\,\in[-0.02,0.01]$ and the effective number of neutrino species \neff$\,\in[2,4]$.

As a fully Bayesian technique, BMA requires us to assign a prior probability to our models; we explore, as a default assumption, a flat or \textit{uniform} prior, consisting of equal prior probability, $P(M_{\rm i})$ in Eq.~\ref{eq:BMA}, to each candidate model. 

\subsection{Dataset}
\label{subsec:data}

In this work, we include the following datasets:
\begin{itemize}
    \item \textbf{CMB}: the full Planck low-$\ell$ and high-$\ell$ likelihood as presented in \cite{Tristram:2024}. This analysis was based on the publicly available codes \textsc{lollipop}\footnote{\hyperlink{https://github.com/planck-npipe/lollipop}{https://github.com/planck-npipe/lollipop}}, covering CMB polarization E and B low-$\ell$ modes in $2\le\ell\le30$, and \textsc{hillipop}\footnote{\hyperlink{https://github.com/planck-npipe/hillipop}{https://github.com/planck-npipe/hillipop}}, that includes the CMB TT, TE and EE multipoles in $30<\ell\le2500$. The TT multipoles in $2\le\ell\le30$ are included through the \textsc{commander} Planck 2018 likelihood \citep{planck2018-V}, based on a gaussianized Blackwell-Rao estimator on the full-sky CMB power spectra from component-spearated CMB maps. 
    \item \textbf{Lensing}: Planck 2018 reconstructed CMB lensing potential power spectrum \citep{planck2018-VIII}.
    \item \textbf{BAO}: Baryon Acoustic Oscillation measurements from SDSS DR7 main galaxy sample \citep{Ross:2014qpa} at $z=0.15$, the updated BAO reconstruction from 6dF galaxy redshift survey \citep{Carter:2018} at $z=0.097$ and the BOSS DR12 \citep{Alam:2016hwk} LOWZ and CMASS galaxy samples at $z=0.38$, $z=0.51$ and $z=0.61$, the eBOSS galaxies and quasars in $0.6\le z\le 2.2$ \citep{Alam:2021}, and the Lyman-$\alpha$ forest samples in $1.8\le z\le 3.5$.
    \item \textbf{BK18}: the Bicep-Keck array CMB polarization data from the 2018 observing season \citep{Ade:2021}.
\end{itemize}

The MCMC runs have been carried out with the publicly available software \textsc{cobaya}\footnote{\hyperlink{https://github.com/CobayaSampler/cobaya}{https://github.com/CobayaSampler/cobaya}} \citep{Torrado:2020dgo,cobaya:ascii}. The resulting chains have been initially analysed with the public code \textsc{getdidst}\footnote{\hyperlink{https://github.com/cmbant/getdist}{https://github.com/cmbant/getdist}} \citep{Lewis:2019xzd}, and then post-processed with \textsc{fast-mpc}\footnote{\hyperlink{https://github.com/simonpara/Fast-MPC}{https://github.com/simonpara/Fast-MPC}} \citep{Paradiso:2023}.

\subsubsection{Choice of estimator}
\begin{figure}
    \centering
    \includegraphics[width=0.6\hsize]{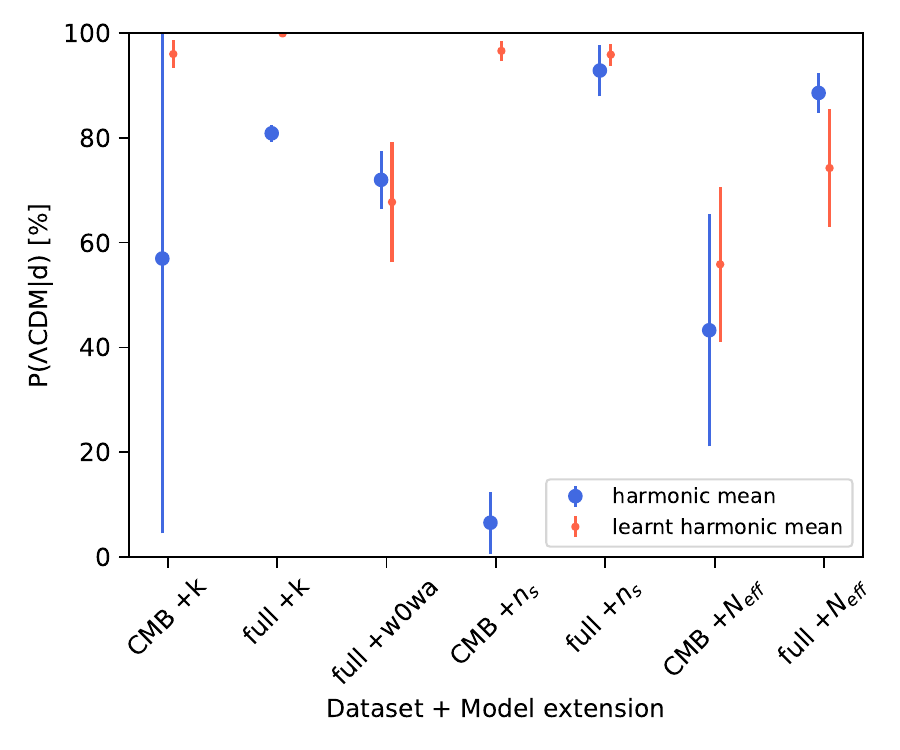}
    \caption{Comparison of P(\LCDM$|d$) between the standard harmonic mean (blue, circle) and the learnt harmonic mean (red, dot) for all the datasets and model extensions configurations examined in this paper: curvature, dark energy EOS, spectral index of primordial fluctuations and effective number of neutrino species. The uncertainties have been computed using a JackKnife splitting for the standard harmonic mean and using the python package built-in function for the learnt harmonic mean estimator.}
    \label{Fig:estimator_comparison}
\end{figure}

In this work, we expand the publicly available tool \textsc{fast-mpc} to exploit either the standard harmonic mean estimator or the learnt harmonic mean one; as discussed, we relied on the already available python package \textsc{Harmonic} for this purpose. We have calculated results for all the considered datasets and model choices using both the estimators. Results comparing the selection probability of the \LCDM\ model to other models are shown in Fig.~\ref{Fig:estimator_comparison}. The model uncertainties were computed via a Jackknife technique for the standard harmonic mean, and used the learnt harmonic mean estimator package otherwise. 
We notice that the two estimators agree within the 67\% error bars in almost every case, with the exception of models with curvature for the full cosmological dataset and models with varying spectral index model case for CMB-only data. Moreover, the learnt harmonic estimator is more precise for these models, whereas the standard harmonic mean suffers from a much higher variance. This can be very dramatic (see for instance the CMB-only, curvature case), and matched the expectation that the harmonic mean can be unstable. We therefore decided to adopt the learnt harmonic mean estimator as the baseline and all results have been computed with this methodology.

%

\subsection{Application 1: Is the Universe flat?}

The first open question about the Cosmological model we address is the one concerning the curvature. While the standard \LCDM\ model is based on a flat Universe with $\Omega_{\rm k}=0$, CMB data showed a tension with this when \Omk\ was included in the analysis of previous Planck data \citep{Ade:2016,planck2018-VI,Rosnberg:2022}. This tension has been drastically reduced to $\lesssim 1\sigma$ with the most recent Planck PR4 data reanalysis from \cite{Tristram:2024}, and removed completely when CMB data are combined with BAO, as shown in the same paper. 

We now present the results of BMA including both the flat \LCDM\ model, and the curved K-\LCDM. For this analysis we considered two dataset configurations: CMB alone and a combination of CMB, CMB lensing, BAO and BK18.
We report in Tab.~\ref{table:curvature_table} the posterior probabilities of the two considered models under a flat model prior assumption and for the two considered datasets. In the flat prior assumption, the CMB data alone suggest a strong model preference for \LCDM\ (86.6\% and 13.4\% for \LCDM\ and K-\LCDM\, respectively), and this preference is even stronger, with a 99.3\% \LCDM\ model probability, when CMB is combined with lensing, BAO and BK18. We show the model posterior probabilities for different model prior choices in Fig.~\ref{fig:priordep} for CMB alone (red line), and for CMB+lens+BAO+BK18 (green line).

\begin{table}[h]
\footnotesize
\centering
\begin{tabular}{lcr}
    \hline
     Dataset & \LCDM & K-\LCDM \\
    \hline 
    \hline
    CMB & $86.6\%$ & $13.4\%$ \\
    CMB+lens+BAO+BK18 & $99.3\%$ & $0.7\%$ \\
    \hline
\end{tabular}
\caption{Model posterior probabilities for the \LCDM\ and K-\LCDM\ under a flat model prior assumption. We compare the CMB dataset alone and including lensing, BAO and Bicep-KECK 2018 results.}
\label{table:curvature_table}
\end{table}

\begin{figure}
    \centering
    \includegraphics[width=0.48\hsize]{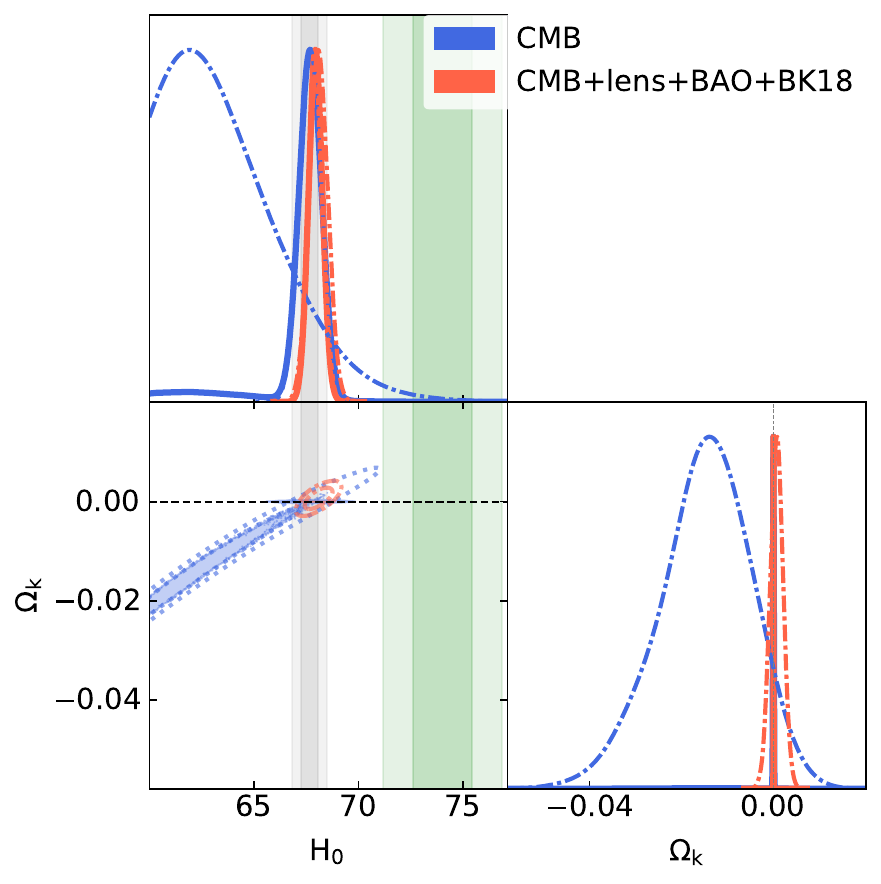}
    \caption{ Posterior distribution in the $\Omega_{\rm k}-\rm{H}_{0}$ parameter space. Solid lines show the model-marginalized constraints with a flat model prior assumption, dotted lines report the \LCDM\ model conditionals and dot-dashed ones the K-\LCDM\ model conditionals.
    We included as vertical bands the the constraint (68\% and 95\% CL) on $H_0$ from Planck 2018 \citep{planck2018-VI} (gray), and from SNIa \citep{Riess:2019cxk} (green), both under the \LCDM\ model assumption.}
    \label{fig:2p-triangle_curvature}
\end{figure}

We present in Fig.~\ref{fig:6lcdm} the marginalized constraints on the 6 \LCDM\ base parameters (red and blue lines); we also show in Fig.~\ref{fig:2p-triangle_curvature} a focus on the the $\Omega_{\rm k}-\rm{H}_{0}$ degeneracy in the BMA context, after marginalizing over the model uncertainty associated to the probabilities of the standard \LCDM\ and the extended K-\LCDM. In the latter figure we notice how a curvature parameter preference for negative values leads to smaller preferred values of the Hubble parameter, in the opposite direction with respect to solving the Hubble tension. The constraint on $\rm H_0$, after marginalizing over model uncertainty, are:
\begin{align}
    &\rm{CMB}  &\rm H_0 = 66.9^{+1.5}_{+0.16} \\
    &\rm{CMB+lens+BAO+BK18}  &\rm H_0 = 67.95\pm 0.34
\end{align}
The CMB estimate is about $0.8\sigma$ below the mean value from the \LCDM-conditional value $67.36\pm 0.61$, and $1.4\sigma$ above the K-\LCDM\ conditional value of $\rm H_0=62.3^{+3.2}_{-3.6}$, as expressed in units of the model marginal estimate's standard deviation. However, the main impact of marginalizing over the model uncertainty is in the increase of the total uncertainty on $\rm H_0$ of $\frac{\sigma_{\rm marg}-\sigma_{\rm cond}}{\sigma_{\rm cond}}\approx 0.36$ but with increased skew towards higher values of the parameter. The CMB+lens+BAO+BK18, on the other hand, is fully consistent with the \LCDM-conditional value $67.95\pm 0.33$, with a $3\%$ increase in the parameter estimate's uncertainty. Similarly, we found the estimates of the curvature density \Omk\ after marginalizing over the model choice to be:
\begin{align}
    &\rm{CMB}  &\Omega_{\rm k} = -0.0023\pm 0.0067 \\
    &\rm{CMB+lens+BAO+BK18}  &\Omega_{\rm k} = 0.00001\pm 0.00019
\end{align}

Both the estimates are fully compatible with a flat Universe within $1\sigma$, despite the uncertainty on $\Omega_{\rm k}$ from CMB alone being $\approx 30\%$ smaller than the K-\LCDM\ conditional estimate $\Omega_{\rm k} = -0.0148^{+0.011}_{-0.0088}$.
The full dataset case also gives a consistent estimate of $\Omega_{\rm k}$ for the K-\LCDM\ model with $\Omega_{\rm k} = 0.0007\pm0014$, with 86\% smaller uncertainty.

\subsection{Application 2: Is the Dark Energy EOS $w=-1$?}

Dark energy contributes ~70\% of the Universe's total energy density, and it is responsible for the accelerated expansion of the cosmos \citep{Riess:1998,Perlmutter:1999}. The parameterization of dark energy equation of state provides a versatile framework for understanding its evolution. In this context, although not related to a physical model, the parameters \wo\ and \wa\ offer valuable insights into the present-day behaviour and the redshift-dependent evolution of dark energy. In this section, we apply BMA to compare and combine three different models: 1) the standard \LCDM, that considers a constant energy density term $\Omega_\Lambda$; 2) a DE EOS with a constant pressure term $w=w_0$; and 3) a time-varying DE EOS $w=w_0+w_{a\rm }(1-a)$. 

We only consider the most constraining dataset including CMB, CMB lensing, BAO and BK18; we did not include also type Ia Supernovae in our analysis, following \cite{planck2018-VI} in order to avoid the $\rm H_0$ tension, and focus our model comparison on DE only. We show in Tab.~\ref{table:deEOS_table} the different model posterior probabilities, with a flat model prior assumption. These show an almost split probability between standard \LCDM\ and the minimal constant EOS \wo\CDM\ model; the most complex \wo\wa\CDM\ is instead discouraged with a relative probability of $0.3\%$.

\begin{table}[h]
\footnotesize
\centering
\begin{tabular}{lccr}
    \hline
     Dataset & \LCDM & \wo\CDM & \wo\wa\CDM \\
    \hline 
    \hline
    CMB+lens+BAO+BK18 & $42.8\%$ & $56.9\%$ & $0.3\%$ \\
    \hline
\end{tabular}
\caption{Model posterior probabilities for the \LCDM\ model, the minimal DE EOS extension \wo\CDM\ and the more complex \wo\wa\CDM\ under a flat model prior assumption.}
\label{table:deEOS_table}
\end{table}

\begin{figure}
    \centering
    \includegraphics[width=0.48\hsize]{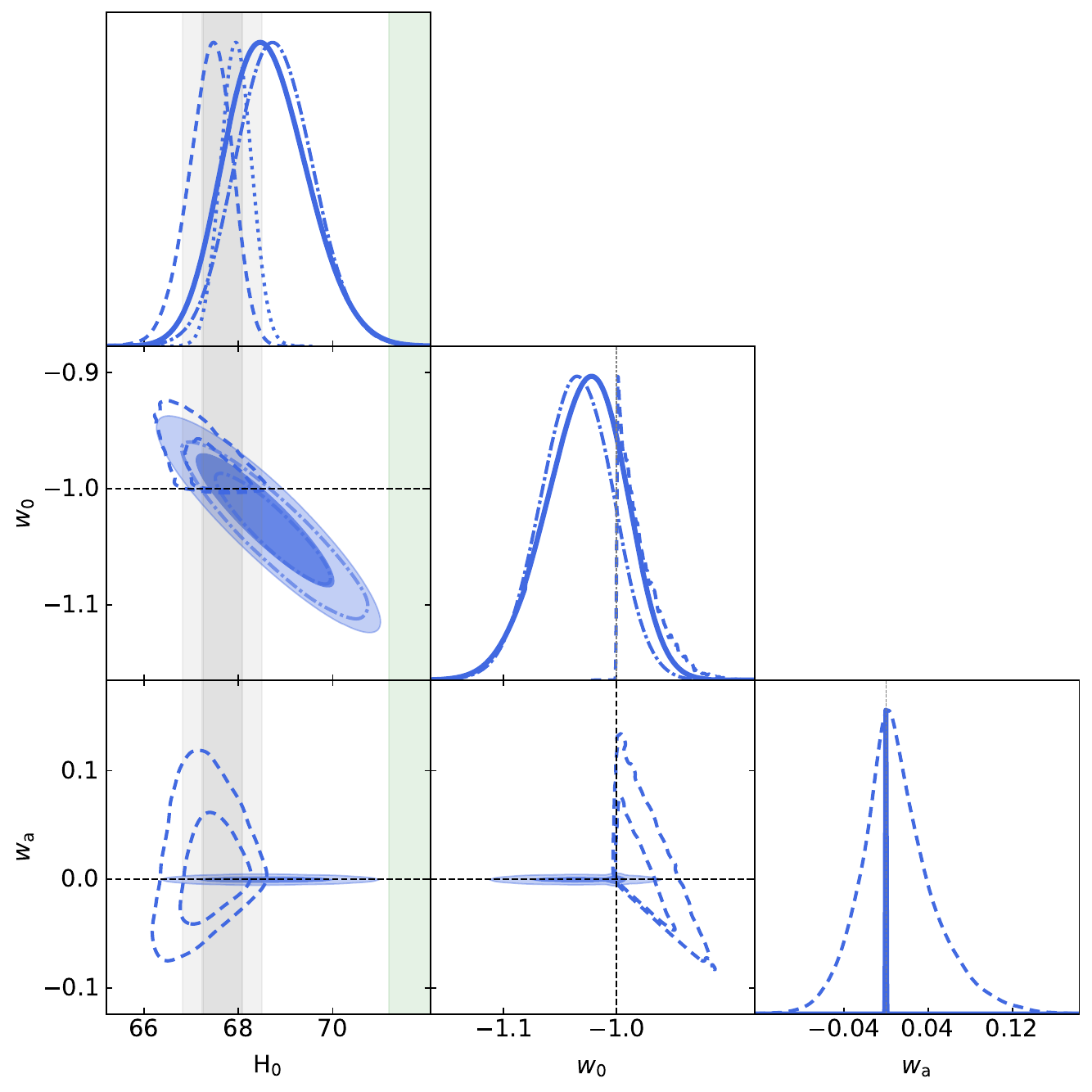}
    \caption{Posterior distributions in the $\rm H_0$-\wo\wa\ parameter space. Solid lines show the model-marginalized constraints with a flat model prior assumption, dotted lines the \LCDM\ ones, dot-dashed the posterior when assuming the \wo\CDM\ model, and dashed lines refer to the assumption of the \wo\wa\CDM\ model.
    We included as vertical (horizontal) bands the the constraint (68\% and 95\% CL) on $H_0$ from Planck 2018 \citep{planck2018-VI} (gray), and from SNIa \citep{Riess:2019cxk} (green), both under the \LCDM\ model assumption.}
    \label{fig:2p-triangle_de}
\end{figure}

Marginalized cosmological parameters are shown in Fig.~\ref{fig:6lcdm} (dark brown lines) for the base 6 parameters, while Fig.~\ref{fig:2p-triangle_de} focuses on the full $\rm H_0$-\wo-\wa\ parameter space. We find that the 68\% credible intervals on these parameters, after including the contribution from the model uncertainty, are:
\begin{align}
    &\rm H_0 = 68.58^{+0.81}_{-0.92} \\
    &w_{\rm 0} = -1.027^{+0.037}_{-0.032} \\
    &w_{\rm a} = 0.0001\pm 0.0028
\end{align}
The model-marginalized posterior distribution of $\rm H_0$ has a higher mean, and a slight skew towards smaller values, However, the higher value of Hubble parameter, together with the $\approx40\%$ larger uncertainty, still do not mitigate the Hubble tension in a significant way. Moreover, the model-marginalized estimates of both \wo\ and \wa\ show a strong compatibility with a constant dark energy EOS with $w_0=-1$.

\subsection{Application 3: Is the Inflationary scalar spectral index $n_s=1$?}

The primordial power spectrum represents the amplitude of density perturbations as a function of their spatial scale. It is generally assumed to have originated from quantum fluctuations during the inflationary epoch (\citep{Guth:1981,Steinhardt:1984}). In this work we focus on the primordial power spectrum index \ns, which describes the scale dependence of the amplitude of density fluctuations. A scale-invariant spectrum would have $n_{\rm s}=1$, meaning that the amplitude of perturbations is the same at all scales. However, CMB observations combined with CMB lensing and BAO showed a slightly lower value of \ns\ in the \LCDM\ model of $n_{\rm s}=0.9665\pm0.0038$ \citep{planck2018-VI}. Additionally, the running of the spectral index, \shortnrun, where $k$ is the wavenumber, is a parameter that characterizes how \ns\ changes with scale. A non-zero \shortnrun\ implies a scale dependence of the spectral index. While the \LCDM\ model assumes a nearly scale-invariant spectrum, the possibility of a running spectral index is still explored in theoretical and observational studies. The current constraints on \shortnrun\ from observations are consistent with zero within the measurement uncertainties \citep{planck2018-VI}.

We explore here three models: 1) the standard \LCDM\ model with \ns\ free to vary in its prior range; 2) a Harrison-Zeldovich primordial power spectrum, corresponding to $n_{\rm s}=1$; and 3) an extended model that has \ns\ free and allows for the running of the spectral index \shortnrun. We computed the relative model posterior probabilities for both a CMB-only dataset and a combination of CMB, CMB lensing, BAO and BK18, and report our results in Tab.~\ref{table:index_table}. Using only CMB data, The Harrison-Zeldovich model is completely discouraged, with a 0\% probability; besides, the \shortnrun\ extension only accounts for a $\sim$6\% probability, and the standard \LCDM\ model with \ns\ free is favoured with a probability of $\sim$93\%. Similarly, when extending the dataset to include CMB lensing, BAO and BK18 we also find the Harrison-Zeldovich model is ruled out, and the standard \LCDM\ model probability dominates on the \shortnrun-\LCDM\ model one with a 90\%-10\% split.

\begin{table}[h]
\footnotesize
\centering
\begin{tabular}{lccr}
    \hline
     Dataset & \LCDM & HZ-\LCDM & \shortnrun-\LCDM \\
    \hline 
    \hline
    CMB & $93.8\%$ & $0\%$ & $6.2\%$\\
    CMB+lens+BAO+BK18 & $89.6\%$ & $0\%$ & $10.4\%$ \\
    \hline
\end{tabular}
\caption{Model posterior probabilities for the \LCDM\ model, the Harrison-Zeldovich HZ-\LCDM\ and the \shortnrun-\LCDM\ under a flat model prior assumption.}
\label{table:index_table}
\end{table}

\begin{figure}
    \centering
    \includegraphics[width=0.48\hsize]{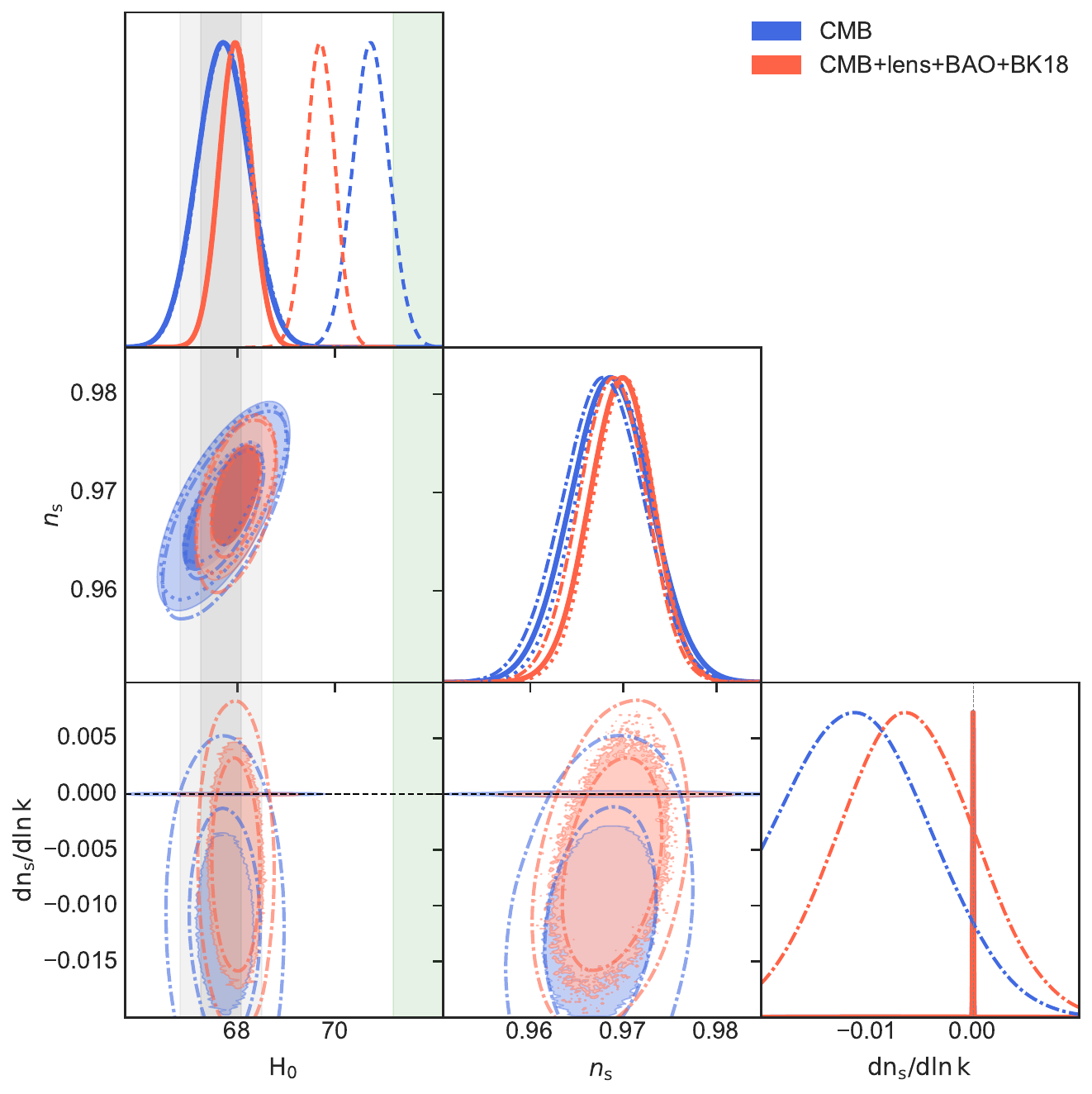}
    \caption{Posterior distribution in the $\rm H_0$-\ns-\shortnrun\ space. Solid lines report the model-marginalized constraints with a flat model-prior assumption, dotted lines the \LCDM\ ones, dashed the HZ-\LCDM\ and dot-dashed lines refer to the \shortnrun-\LCDM\ model.
    The vertical grey band reports the the constraint (68\% and 95\% CL) on $H_0$ from Planck 2018 \citep{planck2018-VI}, and from SNIa \citep{Riess:2019cxk} (green), under the \LCDM\ model assumption.}
    \label{fig:2p-triangle_index}
\end{figure}

The 6 base \LCDM\ parameters marginalized over the model uncertainty are shown in Fig.~\ref{fig:6lcdm} (green lines). We also show in Fig.~\ref{fig:2p-triangle_index} the $\rm H_0$-\ns-\shortnrun\ triangle plot.
Marginalizing over the model uncertainty led to updated estimates of these parameters for the CMB-only case:
\begin{align}
    &\rm H_0 = 67.71\pm 0.53 \\
    &n_{\rm s} = 0.9686\pm 0.0042 \\
    &dn_{\rm s}/\mathrm{d}logk = -0.0010\pm 0.0036
\end{align}
and for the combination of CMB, CMB lensing, BAO and BK18:
\begin{align}
    &\rm H_0 = 67.95\pm 0.33 \\
    &n_{\rm s} = 0.9697\pm 0.0034 \\
    &dn_{\rm s}/\mathrm{d}logk = -0.0015\pm 0.0039
\end{align}
We find a value of $\rm H_0$ only $0.57\sigma$ larger than the \LCDM\ model conditional one in the CMB-only dataset, whereas the full dataset shows a fully compatible estimate with the same uncertainty as in the \LCDM-marginal case from the same dataset. It is worth pointing out two aspects: in the context of the Hubble tension, the HZ model conditional constraint on $\rm H_0$, from CMB data alone, is $\rm H_0=70.73\pm0.38$, approximately $\approx 3\sigma$ away from the SNIa measurement, although this model is  ruled-out by the same data in our analysis, and similarly from the combination of CMB+lensing+BAO+BK18, but with a smaller $\rm H_0$ estimate of $69.70\pm 0.30$. Secondly, our analysis shows a value of \nrun\ compatible with zero in both our datasets, as opposed to the \nrun-\LCDM\ model marginal estimates of $dn_\mathrm{s}/dk = -0.0062\pm 0.0060$ and $dn_\mathrm{s}/dk = -0.0098^{+0.0045}_{-0.0078}$ for the extended dataset and the CMB-only respectively, that point to a $1\sigma$ detection of a negative \nrun.

\subsection{Application 4: How many neutrino species?}

The standard cosmological model assumes a number of relativistic degrees of freedom for neutrinos, \neff$\,\approx 3.046$, slightly larger than 3 to account for the imperfect decoupling at electron-positron annihilation of the neutrino species \citep{Gnedin:1998,Mangano:2005,deSalas:2016}. This parameter can conveniently accommodate any non-interacting, non-decaying, massless particle produced long before recombination; it is defined so that the total relativistic energy density well after electron-positron annihilation is given by:
\begin{equation}
    \rho_{\rm{rad}} = N_{\rm{eff}}\frac{7}{8}\left(\frac{4}{11}\right)^{4/3}\rho_\gamma
\end{equation}
where $\rho_\gamma$ is the photon density. The most updated limits on this quantity comes from a combination of CMB, type Ia Supernovae, BAO and Large Scale Structure data (\cite{DiValentino:2023,DiValentino:2022}.

In this work, we consider CMB alone and the combination of CMB, CMB lensing, BAO and BK18 data to assess the relative model posterior probability of the \LCDM\ standard model, and the \neff-\LCDM\ model that accounts for additional relativistic species. We report the model posterior probabilities for the two datasets in Tab.~\ref{table:neutrino_table} for a flat model prior, and illustrate the prior dependence in Fig.~\ref{fig:priordep} for the CMB alone (blue line) and CMB+lens+BAO+BK18 (purple line).  For both datasets, we find no preference for either the \LCDM\ or the \neff-\LCDM\ model with a split probability of 53.3\%-46.5\% and 56\%-44\% for CMB alone and in combination with lensing, BAO and BK18, respectively.

\begin{table}[h]
\footnotesize
\centering
\begin{tabular}{lcr}
    \hline
     Dataset & \LCDM & \neff-\LCDM \\
    \hline 
    \hline
    CMB & $53.5\%$ & $46.5\%$ \\
    CMB+lens+BAO+BK18 & $56.0\%$ & $44.0\%$ \\
    \hline
\end{tabular}
\caption{Model posterior probabilities for the \LCDM\ and \neff-\LCDM\ under a flat model prior assumption. We compare the CMB dataset alone and including lensing, BAO and Bicep-KECK 2018 results.}
\label{table:neutrino_table}
\end{table}

\begin{figure}
    \centering
    \includegraphics[width=0.48\hsize]{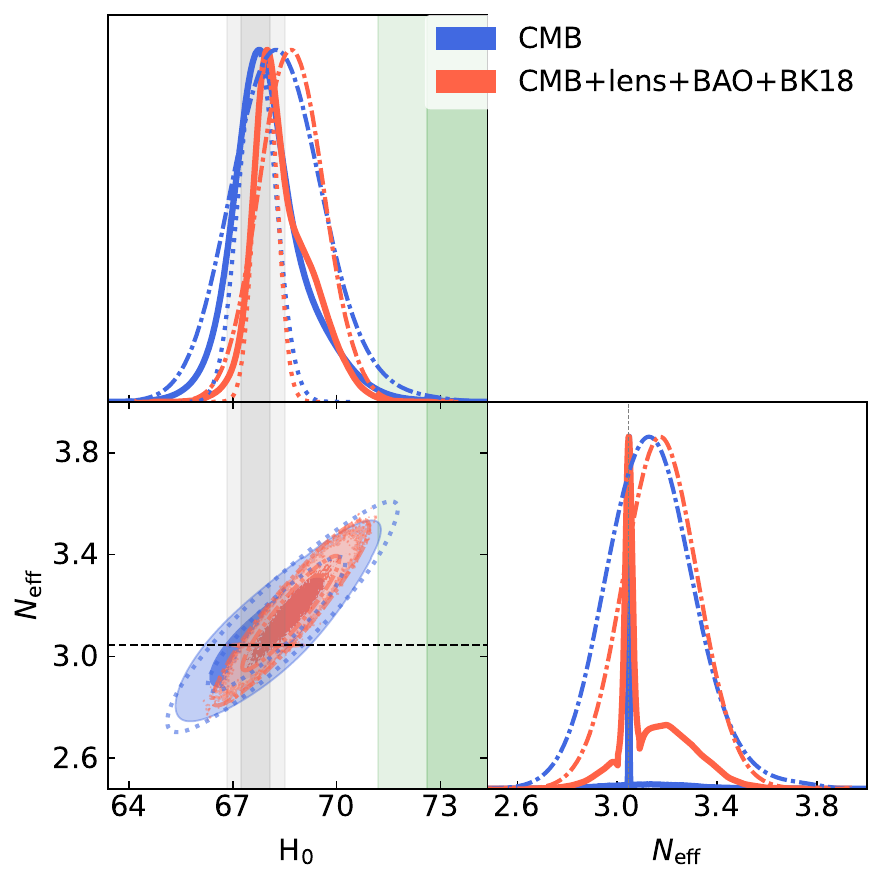}
    \caption{Posterior distributions in the $\rm H_0$-\neff parameters space; solid lines report the model-marginalized constraints with a flat model prior assumption, dotted lines refer to \LCDM\ model-conditional ones and dash-dotted lines show the \neff-\LCDM\ model-conditional constraints.
    Model posterior probabilities for the \LCDM\ model against the \neff-\LCDM\ model under a flat model prior assumption. We included as vertical bands the the constraint (68\% and 95\% CL) on $H_0$ from Planck 2018 \citep{planck2018-VI} (gray), and from SNIa \citep{Riess:2019cxk} (green), both under the \LCDM\ model assumption.}
    \label{fig:H0-neff}
\end{figure}

\begin{figure}
    \centering
    \includegraphics[width=1\hsize]{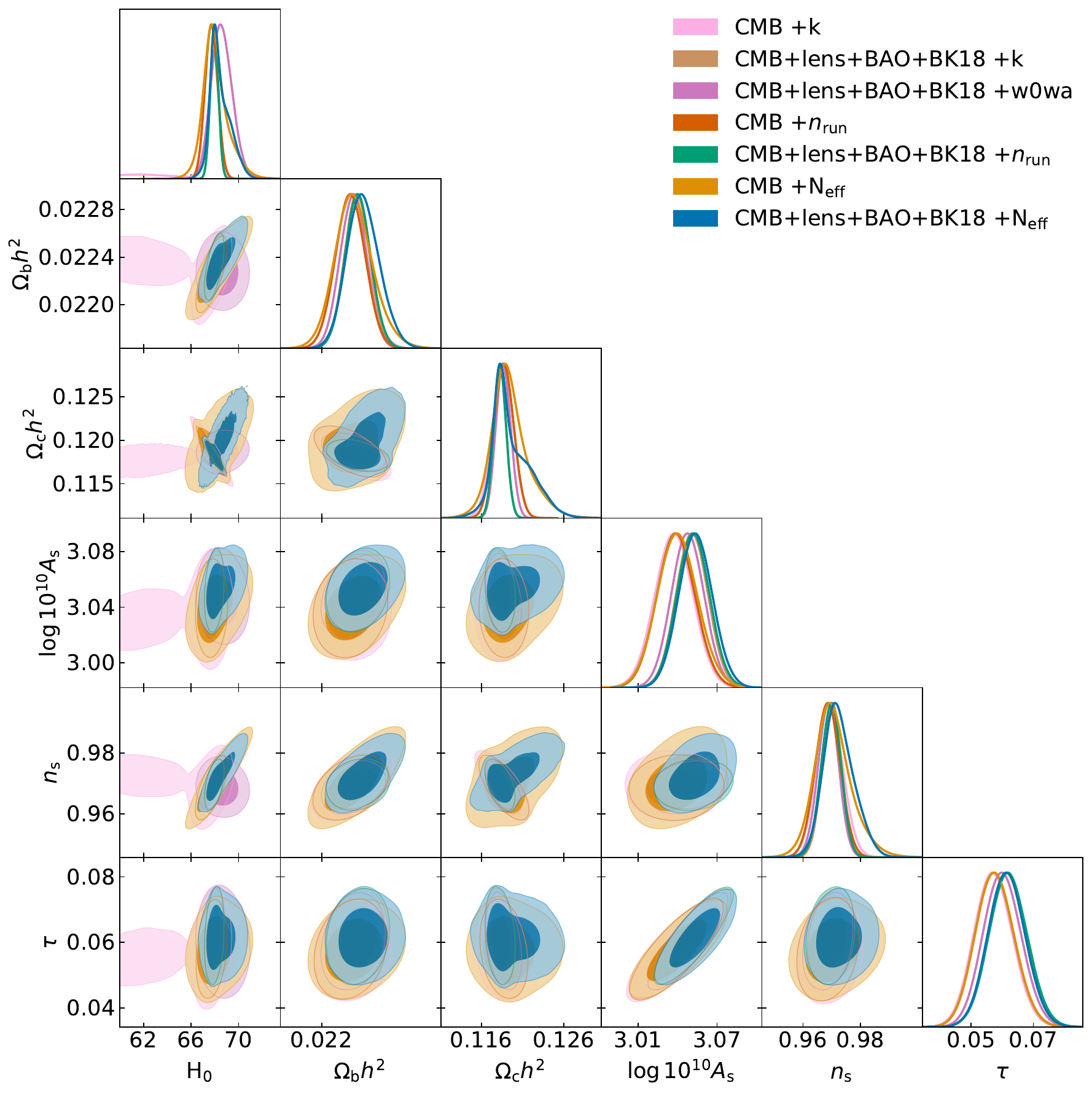}
    \caption{The 6 \LCDM\ base parameters as marginalized over the model uncertainty, for the datasets and \LCDM\ extension considered in this work: CMB only and full CMB+lensing+BAO+BK18 datasets; the \LCDM\ is compared to curvature, dark energy EOS (constant and tyme-varying), scale-invariant power spectrum and \shortnrun, and additional ultra-relativistic species at decoupling through the parameter \neff.}
    \label{fig:6lcdm}
\end{figure}

As for the other model comparisons in this paper, we propagate the model uncertainties to the 6 base parameters of the \LCDM\ model common to all its extensions, and show the marginalized posterior distributions as orange and blue lines in Fig.~\ref{fig:6lcdm}. We focus on the $\rm H_0$-\neff parameters space in Fig.~\ref{fig:H0-neff}. We find the model-marginalized constraints on $\rm H_0$:
\begin{align}
    &\rm{CMB}  & \rm H_0 = 68.05^{+0.75}_{-1.2} \\
    &\rm{CMB+lens+BAO+BK18}  & \rm H_0 = 68.44^{+0.61}_{-0.97}
\end{align}

and similarly, for $N_{\rm eff}$:
\begin{align}
    &\rm{CMB}  & N_{\rm{eff}} = 3.10^{+0.14}_{-0.060} \\
    &\rm{CMB+lens+BAO+BK18}  & N_{\rm{eff}} = 3.13\pm 0.13
\end{align}

For both the datasets, the marginalization over the model uncertainty when considering additional relativistic degrees of freedom through \neff\ drives $\rm H_0$ in the right direction to solve the Hubble tension. Specifically, we find that the CMB data alone give a $1.1\sigma$ higher Hubble parameter estimate, with a $\approx60\%$ larger uncertainty as compared to the \LCDM\ model-conditional value. For the more extensive dataset, we  recover a value of $\rm H_0$ slightly higher, $0.8\sigma$ above the \LCDM-conditional one and with a standard deviation $140\%$ larger than the \LCDM-conditional value for the same dataset, and $0.26\sigma$ lower and with a standard deviation 15\% smaller than the \neff-\LCDM-conditional case. Those results reflect the poor constraint we have on the model from the considered datasets, and the final parameter estimate is effectively an average of those from the two models.

\section{Conclusions}
\label{sec:conclusions}

\begin{figure}
    \centering
    \includegraphics[width=0.7\hsize]{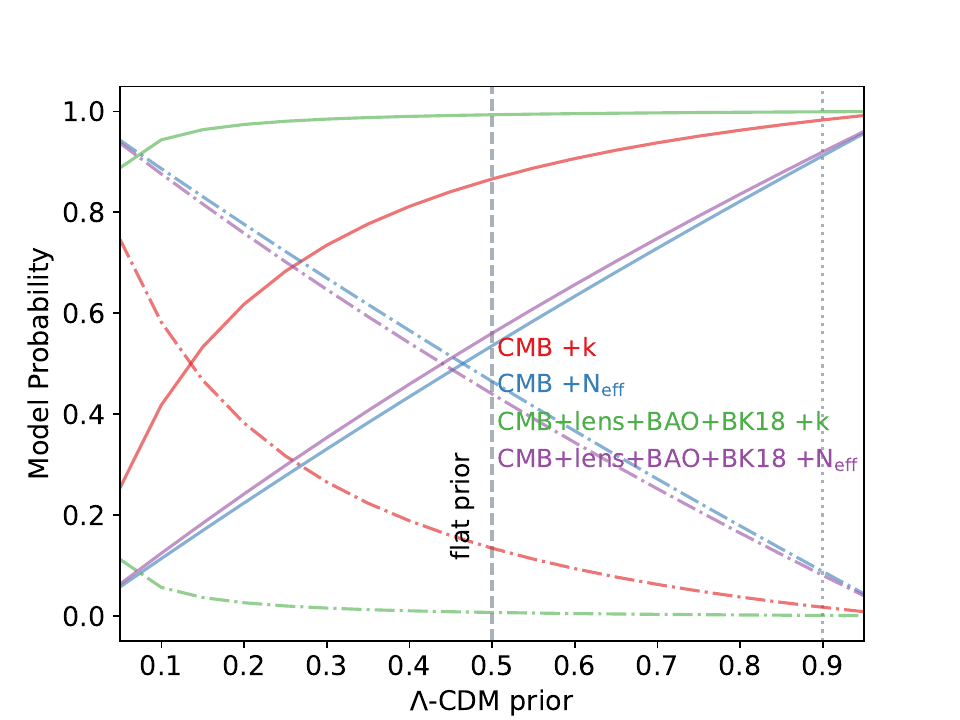}
    \caption{Prior dependence of the model posterior probabilities for the two-models problems considered in this work: the model posterior probability for the comparison \LCDM\ vs K-\LCDM\ for CMB only and the full dataset respectively; the model posterior probabilities relative to the comparison \LCDM\ vs \neff-\LCDM\, for CMB only and the full dataset respectively. The solid lines refer to $P(\Lambda\rm{CDM})$, whereas dotted-dashed lines report the K-\LCDM\ and \neff-\LCDM\ model posterior probabilities ($1-P(\Lambda\rm{CDM})$). The vertical black dashed and dotted lines report the split prior, and a more extreme $90\%:10\%$ allocation in favour of \LCDM, respectively.}
    \label{fig:priordep}
\end{figure}

In this paper we have presented an updated review, following the spirit of previous works \cite{Parkinsons:2013,Paradiso:2023}, addressing current questions in Cosmology through Bayesian model averaging and selection. This methodology provided us with a principled approach to assign posterior probabilities to different models in a Bayesian context, and to seamlessly propagate this additional source of uncertainty all the way to the estimated cosmological parameters. In other words, we let the data express a preference---or lack thereof---between different models. The analysis has been carried out exploiting the importance sampling approach presented in \cite{Paradiso:2023} (\textsc{Fast-MPC}), which we have improved by adopting the much stabler learnt harmonic mean estimator in place of the standard harmonic mean.

Firstly, we addressed the curvature problem, by considering the flat \LCDM\ model and the curved K-\LCDM. We found that both the CMB data alone and in combination with CMB lensing, BAO and BK 2019 express a strong preference for the standard \LCDM\ model with a probabilty above 80\% and 90\% respectively.

We then considered three models for the evolution of Dark Energy: the standard \LCDM\ with a cosmological constant $\Omega_\Lambda\approx0.7$, a DE component with a constant EOS $w=w_0$ and a DE component with a time-varying EOS $w=w_0+(1-a)w_{\rm a}$. We found that the combination of CMB, lensing, BAO and BK 2018 data, under a flat model prior, is not able to express a strong preference between the standard \LCDM\ and the \wo\CDM\ model, with a split probability of 43\%-57\%. The more complex \wo\wa\CDM\ is disfavoured with only a $9.1\%$ probability.

For the inflationary scalar spectral index, we considered three models: the standard \LCDM\ model with varying \ns, a Harrison-Zeldovich spectrum, corresponding to a fixed $n_{\rm s}=1$ and an extended model with varying \ns\ and running of the spectral index \shortnrun. We considered the CMB data alone and in combination with CMB lensing, BAO and BK 2018 for this problem, and found a very high preference for the \LCDM\ model, with a probability of ~90\% when a flat model prior is assumed, for both the datasets.

Finally, we investigated the possibility of having additional massless hot-dark matter species in the Universe by varying the \neff\ parameter. We again found that both the CMB alone and the extendend CMB+Lensing+BAO+BK18 dataset exhibit a split model probability between \neff-\LCDM\ and \LCDM, with a \LCDM\ model probability of 54.5\% and 56\% for CMB only and extended dataset respectively (with a flat model prior). 

We have tested the impact of different model prior choices in the two-model applications, namely the curvature and the \neff-\LCDM\ extensions. We show in Fig.~\ref{fig:priordep} the model posterior probability dependence from the initial model prior choice (solid: \LCDM, dot-dashed: K-\LCDM, \neff-\LCDM), when considering different datasets. While the priors are obviously important for our inferences, we see that our conclusions for curvature are relatively robust to a range of prior probabilities around the flat value, whereas the model probabilities are dominated by the prior choice for the neutrino extension scenario, due to the poor constraining power of the considered datasets. We expect a significant improvement on the latter model constraint from future CMB and LSS surveys, as smaller angular scales become accessible shedding light on the role of neutrinos on the growth of cosmic structures.

We also assessed the impact of our model uncertainty marginalization through BMA on the Hubble tension, and found, consistent with current literature, that the model uncertainty does not solve the discrepancy in the Hubble parameter estimate from CMB and BAO versus SNIa. However, we highlight the role BMA could play in discriminating between models in terms of the preference expressed by data themselves.

In conclusion, our work has used an improvement of the method of \cite{Paradiso:2023} to address four interesting model choices within cosmology. Considering decisions between 2 or 3 models, we have let the data tell us whether any extensions to \LCDM\ are required. 
Our analysis confirms the standard \LCDM\ model as a very complete description of the Universe. The work also shows how current data are not able to discriminate between a some \LCDM\ extensions, including the DE EOS and the number of relativistic species, for which we expect major improvements from future data, particularly with the release of results from stage-IV Dark Energy experiments including the Dark Energy Spectroscopic Instrument (DESI; \cite{desi:2016}) and the Euclid satellite mission \cite{Laureijs:2011,euclid_I:2022}.


\acknowledgments

All authors acknowledge support from the Canadian Government through a New Frontiers in Research Fund (NFRF) Exploration grant.

WP acknowledges the support of the Natural Sciences and Engineering Research Council of Canada (NSERC), [funding reference number RGPIN-2019-03908] and from the Canadian Space Agency.

GM acknowledges the support of the Natural Sciences and Engineering Research Council of Canada (NSERC), [RGPIN-2022-03068 and DGECR-2022-004433].

Research at Perimeter Institute is supported in part by the Government of Canada through the Department of Innovation, Science and Economic Development Canada and by the Province of Ontario through the Ministry of Colleges and Universities.

This research was enabled in part by support provided by Compute Ontario (\hyperlink{computeontario.ca}{computeontario}) and the Digital Research Alliance of Canada (\hyperlink{alliancecan.ca}{alliancecan}).



\bibliographystyle{JHEP}
\bibliography{biblio.bib}

\end{document}